\documentclass[twocolumn,aps,pra,amsmath,amssymb,showpacs,superscriptaddress]{revtex4}

\usepackage{graphicx}
\usepackage{graphics}
\usepackage[latin1]{inputenc}
\usepackage{latexsym}
\usepackage{amsmath}
\usepackage{amssymb}

\def\endproof{\vrule height6pt width6pt depth0pt}

\begin{document}
\title{Minimum detection efficiency for a loophole-free violation of
 the Braunstein-Caves chained Bell inequalities}
\author{Ad{\'a}n Cabello}
 \email{adan@us.es}
 \affiliation{Departamento de F{\'\i}sica Aplicada II, Universidad
 de Sevilla, E-41012 Sevilla, Spain}
\author{Jan-{\AA}ke Larsson}
%\email{jalar@mai.liu.se}
\affiliation{Matematiska Institutionen, Link\"opings Universitet,
  SE-58183 Link\"oping, Sweden} \author{David Rodr{\'\i}guez}
%\email{davidroba@hotmail.com}
\affiliation{Departamento de F{\'\i}sica Aplicada II, Universidad de
  Sevilla, E-41012 Sevilla, Spain}

%%%%%%%%%%%%%%%%%%%%%%%%%%%%%%%%%%%%%%%%%%%%%%%%%%%%%%%%%%%%%%%%%%%

\date{\today}
%First version: 26 November 2007
%This version: 10 June 2009 (Vienna, after PRA proofs)

%%%%%%%%%%%%%%%%%%%%%%%%%%%%%%%%%%%%%%%%%%%%%%%%%%%%%%%%%%%%%%%%%%%

\begin{abstract}
The chained Bell inequalities of Braunstein and Caves involving $N$
settings per observer have some interesting applications. Here we
obtain the minimum detection efficiency required for a loophole-free
violation of the Braunstein-Caves inequalities for any $N \ge 2$. We
discuss both the case in which both particles are detected with the
same efficiency and the case in which the particles are detected
with different efficiencies.
\end{abstract}

%%%%%%%%%%%%%%%%%%%%%%%%%%%%%%%%%%%%%%%%%%%%%%%%%%%%%%%%%%%%%%%%%%%

\pacs{03.65.Ud,
%Entanglement and quantum nonlocality
%(e.g. EPR paradox, Bell's inequalities, GHZ states, etc.)
  03.67.Mn,
%Entanglement production, characterization and manipulation,
  03.67.-a,
%Quantum information
  42.50.Xa}
%Optical tests of quantum theory

\maketitle

%%%%%%%%%%%%%%%%%%%%%%%%%%%%%%%%%%%%%%%%%%%%%%%%%%%%%%%%%%%%%%%%%%%

\section{Introduction}

%%%%%%%%%%%%%%%%%%%%%%%%%%%%%%%%%%%%%%%%%%%%%%%%%%%%%%%%%%%%%%%%%%%

Soon after the Clauser-Horne-Shimony-Holt (CHSH) generalization
\cite{CHSH69} of the original Bell inequality \cite{Bell64}, Wigner
\cite{Wigner70} and Pearle \cite{Pearle70} realized that it is
possible to make a local hidden variable (LHV) model which produces
predictions in agreement with the predictions of quantum mechanics
(QM) for a maximal violation of the CHSH Bell inequality, if each
particle has not two, but three possible responses to the local
measurements: being detected by the detector labeled $-1$, being
detected by the detector labeled $+1$, or being undetected. ``Then
instead of four possible outcomes (\ldots), there are nine possible
outcomes. In one of these outcomes, neither particle is detected
(\ldots). In four of these outcomes one of the particles is not
detected. If the experimenter rejects these data (in the belief that
the apparatus is not functioning properly and that if it had been
functioning properly, the data recorded would have been
representative of the accepted data) (\ldots), it is possible to
produce a local hidden variable theory [which gives] predictions in
agreement with the predictions of quantum theory'' \cite{Pearle70}.

This is the origin of the so-called detection loophole of
experimental tests of the violation of Bell inequalities. In most
experimental ``violations'' of Bell inequalities, the overall
detector efficiency (defined as the ratio of detected to produced
particles) is below $0.2$ (two remarkable exceptions are
\cite{RKVSIMW01, MMMOM08}, where it is almost $1$), and the
experimenter rejects all the events where at least one of the
particles is not detected, and assumes that the remaining data is
representative of the data recorded had the efficiency of the
detectors been perfect (this is the so-called fair-sampling
assumption). This is an auxiliary assumption that restricts the
studied class of LHV models considerably.

This paper focuses on the question of what minimum overall detection
efficiency $\eta_{\rm crit}$ is required to escape from the
detection loophole. In other words, how good our detectors need to
be to give a conclusive experimental violation of a Bell inequality
without the fair-sampling assumption. The bound $\eta_{\rm crit}$ is
the value of the detected- to emitted-particle ratio such that, if
$\eta\le\eta_{\rm{crit}}$, there is an LHV model reproducing the
predictions of QM, but no such LHV models exist if $\eta > \eta_{\rm
crit}$.

For the CHSH Bell inequality, and assuming a perfect preparation,
$\eta_{\rm crit}=2(\sqrt{2}-1)\approx 0.83$ if all particles have
detected with the same efficiency \cite{GM87, Larsson98}, and
$\eta_{\rm crit}=1/\sqrt2\approx0.71$ if one of the particles is
always detected \cite{CL07, BGSS07}.

Mermin proposed an $n$-party two-setting generalization of the
(two-party two-setting) CHSH Bell inequality \cite{Mermin90}. For
the Mermin Bell inequalities, it has been recently proven that
$\eta_{\rm{crit}}(n)=n/(2N-2)$ \cite{CRV08}. The amount of violation
${\cal D}$ (defined, for Bell inequalities involving only averages
of products of local operators, as the ratio between the quantum
prediction and the bound of the Bell inequality) grows with the
number of parties $n$ as ${\cal D}(n)=2^{(N-1)/2}$. Therefore, for
the Mermin Bell inequalities,
$\eta_{\rm{crit}}=[2+(\log2/\log{\cal{D}})]/4$; it seems likely that
there is a close relation between $\eta_{\rm crit}$ and ${\cal{D}}$
for other generalizations as well.

Braunstein and Caves (BC) proposed a two-party $N$-setting
generalization of the CHSH Bell inequality \cite{BC89, BC90}, in
which the first observer can choose one out of $N$ alternative
experiments $A_1$, $A_3$, \ldots, $A_{2N-1}$, and the second
observer one out of $N$ alternative experiments $B_2$, $B_4$,
\ldots, $B_{2N}$, each of them having only outcomes $+1$ or $-1$.
The BC chained Bell inequalities (in the case of ideal detectors)
are
\begin{equation}
  \begin{split}
    |&E( A_1 B_2 ) + E( A_3 B_2 ) + E( A_3 B_4 ) + E( A_5 B_4 ) + \cdots
    \\  & + E( A_{2N-1} B_{2N} ) - E( A_1 B_{2N}) | \le 2 N-2.
  \end{split}
\label{BC}
\end{equation}
These inequalities are violated by correlations
$\langle{}A_iB_j\rangle$ obtained from QM. For instance
\cite{Peres93}, for the state
\begin{equation}
|\psi^-\rangle = \frac{1}{\sqrt{2}}
\left(|01\rangle-|10\rangle\right), \label{singlete}
\end{equation}
choosing
\begin{subequations}
\begin{align}
  A_j = \cos (j \pi/2N) \sigma_x + \sin (j \pi/2N) \sigma_z,\\
  B_k = \cos (k \pi/2N) \sigma_x + \sin (k \pi/2N) \sigma_z,
\end{align}
\label{localobservables}
\end{subequations}
we obtain
\begin{equation}
 \begin{split}
 \langle A_1 B_2 \rangle &= \langle A_3 B_2 \rangle = \langle A_3 B_4
 \rangle
 \\
 &= \cdots = \langle A_{2N-1} B_{2N} \rangle
 \\
 &= - \langle A_1 B_{2N} \rangle \\
 &= \cos(\pi/2N). \label{averages}
 \end{split}
\end{equation}
Therefore, the violation is
\begin{equation}
{\cal D}(N) = \frac{2N \cos(\pi/2N)}{2N-2}. \label{DBC}
\end{equation}
That is, ${\cal D}(2)=\sqrt{2}\approx 1.414$ (which is the maximum
possible violation of the CHSH Bell inequality in QM
\cite{Tsirelson80}) and ${\cal D}(3)=3 \sqrt{3}/4 \approx 1.299$.
The violation decreases with $n$. Indeed, Eq.~(\ref{DBC}) gives the
maximum possible violation of the BC chained Bell inequalities
(\ref{BC}) in QM \cite{Wehner06}.

Violations of the BC chained Bell inequalities have been observed
(under the fair-sampling assumption) using pairs of photons
entangled in polarization, with $N=3$, $4$ \cite{BBDH97}, and even
$N=21$ settings per observer \cite{BDDM05}.

The BC chained Bell inequalities have some interesting applications
in situations where the CHSH Bell inequality is inadequate. For
instance, the use of a BC inequality with $N=3$ solves a problem in
Franson's CHSH Bell experiment \cite{AKLZ99}, and reduces the number
of trials needed to rule out local realism in experiments with
perfect detection efficiency \cite{Peres00}. Moreover, the use of BC
inequalities with higher values of $N$ improves the security of
quantum key distribution protocols \cite{BHK05}, and has been also
used to investigate nonlocal theories \cite{BKP06, CR08}.

The aim of this paper is to calculate $\eta_{\rm crit}(N)$ for the
maximum possible violation of the BC chained Bell inequalities
(\ref{BC}) assuming a perfect preparation.

In Sec.~\ref{2A} we introduce some definitions. In Sec.~\ref{2B} we
state the main result. The necessary condition is proven in
Sec.~\ref{2C}. Both the case with equal (symmetric) and unequal
(asymmetric) efficiencies for both particles are discussed. To prove
the sufficient condition, explicit LHV models are built for both
cases. The sufficient conditions for symmetric and asymmetric
efficiencies are developed in Secs.~\ref{2D} and \ref{2E},
respectively.

In Sec.~\ref{3} we present the conclusions and discuss the relation
between the amount of violation ${\cal D}$ and $\eta_{\rm crit}$ for
the BC inequalities and the effect of non-perfect visibilities in
the state preparation.

%%%%%%%%%%%%%%%%%%%%%%%%%%%%%%%%%%%%%%%%%%%%%%%%%%%%%%%%%%%%%%%%%%%

\section{Detection efficiency for the Braunstein-Caves chained Bell inequality}

%%%%%%%%%%%%%%%%%%%%%%%%%%%%%%%%%%%%%%%%%%%%%%%%%%%%%%%%%%%%%%%%%%%

\subsection{\label{2A}Basic definitions}

%%%%%%%%%%%%%%%%%%%%%%%%%%%%%%%%%%%%%%%%%%%%%%%%%%%%%%%%%%%%%%%%%%%

In an LHV model, the result of a measurement of $A_j$ on particle 1
and $B_k$ on particle 2 is predetermined. This information can be
summarized in the {\em state of LHV of an individual pair of
particles} (hereafter simply called state), which is a list
$\{A_1,A_3,\ldots,A_{2N-1};B_2,B_4,\ldots,B_{2N}\}$ of $2N$ {\em
instructions}. For a given measurement $A_j$ (or $B_k$), the
possible instructions are: ``give a detection in the detector
$-1$,'' ``give a detection in the detector $+1$,'' and ``do not give
a detection.'' We will denote these instructions as $-1$, $+1$, and
$0$, respectively. Therefore, each state is represented by a list of
$2N$ values in $\{-1,+1,0\}$.

Because of the special status of the value 0 (``no detection'') it is
not easy to estimate $E(A_jB_k)$ from experiment. An estimate would
need counting the number of ``no detection'' events that has occurred,
and this is a nontrivial exercise. The usual approach is to delete (or
rather, disregard) the ``no detection'' events and calculate the
conditional correlation, given that a coincidence has occurred. We will
use the notation $\Lambda_{A_jB_k}$ for the ensemble of pairs that
give rise to a coincidence, i.e., the ensemble where $A_j\neq0$ and
$B_k\neq0$.

Using this notation, the averages easily obtainable from experiments
are conditional averages on the form $E(A_j|\Lambda_{A_j})$, and
similarly conditional correlations on the form
$E(A_jB_k|\Lambda_{A_jB_k})$, both averages over obtained data. In
general, given an ensemble $\Lambda$ of pairs, $E(A_j|\Lambda)$ will
denote the average restricted to $\Lambda$. If we divide the ensemble
$\Lambda$ into disjoint subensembles $\Lambda_i$,
\begin{equation}
  E(A_jB_k|\Lambda)
  =\sum_{i} E(A_jB_k|\Lambda_i) P(\Lambda_i|\Lambda). \label{average_product}
\end{equation}

An {\em LHV model} for a given Bell experiment is an ensemble of
pairs, each of them with its own state, which satisfies the
predictions of QM for that experiment and reproduces the behavior of
actual detectors. For example, in order to reproduce the predictions
of QM for state (\ref{singlete}) and local observables
(\ref{localobservables}), the LHV model must satisfy
\begin{subequations}
\begin{align}
E( A_j |\Lambda_{A_j})& = 0,\;\forall j\in\{1,3,\ldots, 2N-1\},
\label{zero_mean_1}
\\
E( B_k |\Lambda_{B_k})& = 0,\;\forall k\in \{2,4,\ldots, 2N\},
\label{zero_mean_2}
\end{align}
\end{subequations}
and also must satisfy [from Eqs.~(\ref{averages})]
\begin{equation}
  \begin{split}
    E(A_1B_2|\Lambda_{A_1B_2})&=E(A_3B_2|\Lambda_{A_3B_2})\\
    &=E(A_3B_4|\Lambda_{A_3B_4})\\
    &=\cdots = E(A_{2N-1}B_{2N}|\Lambda_{A_{2N-1}B_{2N}})\\
    &=-E(A_1B_{2N}|\Lambda_{A_1B_{2N}})\\
    &=\cos(\pi/2N).\label{eq:1}
  \end{split}
\end{equation}

From our LHV model, we can now obtain probabilities like
$P(\Lambda_{A_j})$, the probability that $A_j$ is nonzero,
$P(\Lambda_{A_jB_k})$, the probability that both $A_j$ and $B_k$ are
nonzero, and $P(\Lambda_{A_j}|\Lambda_{B_k})$, the conditional
probability that $A_j\neq0$ given that $B_k\neq0$. Note that the
last probability is simple to extract from an experiment while the
former two are more difficult to get at. Also,
$P(\Lambda_{A_jB_k})=P(\Lambda_{A_j}|\Lambda_{B_k})P(\Lambda_{B_k})$.
We will use the minimum conditional detection probability as an
efficiency measure of our setups. In general, this means that the
two detection sites can have individual efficiency measures,
\begin{subequations}
  \begin{align}
    \eta_{A}&\stackrel{\textup{def}}=
    \min_{j,k}
    P(\Lambda_{A_{j}}|\Lambda_{B_{k}}),
    \label{heta_A}\\
    \eta_{B}&\stackrel{\textup{def}}=
    \min_{j,k}
    P(\Lambda_{B_{k}}|\Lambda_{A_{j}}).
    \label{heta_B}
  \end{align}
\end{subequations}
The efficiency of the whole setup can be measured as
\begin{equation}
  \label{eq:27}
      \eta\stackrel{\textup{def}}=
    \min \eta_A,\eta_B.
\end{equation}

In order to reproduce the behavior of actual detectors, we will
construct the LHV model to give nondetections at a constant
probability (independent of measurement settings) that are
statistically independent between the two sites. This may seem like
a severe restriction on the model, but as we will see, the model
will be capable of reaching the bound $\eta_{\rm{crit}}$ even with
this restriction. In the model, this corresponds to that the
probabilities must satisfy
\begin{subequations}
\begin{align}
&P(\Lambda_{A_j})=\eta_{A},
\label{eff_1}\\
&P(\Lambda_{B_k})=\eta_{B},
\label{eff_2}\\
&P(\Lambda_{A_j}|\Lambda_{B_k})=\eta_{A},
\label{eff_3}\\
&P(\Lambda_{B_k}|\Lambda_{A_j})=\eta_{B}, \label{eff_4}
\end{align}
\label{eff}
\end{subequations}
for the relevant combinations of $j\in\{1,3,\ldots, 2N-1\}$, and
$k\in\{2,4,\ldots, 2N\}$.

%%%%%%%%%%%%%%%%%%%%%%%%%%%%%%%%%%%%%%%%%%%%%%%%%%%%%%%%%%%%%%%%%%%

For our purposes, it is also useful to realize that any LHV model can
also be defined as a set of states and their probabilities of
appearance.  Clearly, the same applies to any of its subensembles (a
specific value of the LHV is just a particular kind of ensemble), and,
by definition, those probabilities will always be relative to the
whole LHV model. This choice makes their interpretation as
probabilities consistent on the probability space defined by the LHV
model.

%%%%%%%%%%%%%%%%%%%%%%%%%%%%%%%%%%%%%%%%%%%%%%%%%%%%%%%%%%%%%%%%%%%

\subsection{\label{2B}Main results}

%%%%%%%%%%%%%%%%%%%%%%%%%%%%%%%%%%%%%%%%%%%%%%%%%%%%%%%%%%%%%%%%%%%

In what follows, we will prove the following theorem:

{\em Theorem 1.} The BC inequality (\ref{BC}) has a well-defined
critical efficiency. That is, an efficiency below or equal to this
critical value is necessary and sufficient for the existence of an
LHV model giving the quantum violation of the inequality. Moreover,
the value in the symmetric case ($\eta_A=\eta_B=\eta$) is
\begin{equation}
\eta_{\rm crit}(N) =
\frac{2}{\frac{N}{N-1}\cos\left(\frac{\pi}{2N}\right)+1},
\label{heta_quantum_sim}
\end{equation}
and, when $\eta_{A}\neq\eta_{B}$, the relation between $\eta_{A\;\rm
crit}$ and $\eta_{B\;\rm crit}$ is
\begin{equation}
  \eta_{A\;\rm crit}(N)
  =
  \frac{1}{\frac{N}{N-1}\cos\left(\frac{\pi}{2N}\right)+1
  -\frac1{\eta_{B\;\rm crit}}}.
\label{heta_quantum_asim}
\end{equation}

%%%%%%%%%%%%%%%%%%%%%%%%%%%%%%%%%%%%%%%%%%%%%%%%%%%%%%%%%%%%%%%%%%%

\subsection{\label{2C}Necessary condition}

%%%%%%%%%%%%%%%%%%%%%%%%%%%%%%%%%%%%%%%%%%%%%%%%%%%%%%%%%%%%%%%%%%%

We now prove that the right-hand sides of
Eqs.~(\ref{heta_quantum_sim}) and (\ref{heta_quantum_asim}) are
indeed upper bounds. The following proof does not need to assume
independent errors [e.g., that
$P(\Lambda_{A_j}|\Lambda_{B_k})=P(\Lambda_{A_j})$] or constant error
rates [e.g., that $P(\Lambda_{A_j})=P(\Lambda_{A_k})$], hinted at
above.

In the ideal case, the BC inequalities assert
\begin{equation}
  \begin{split}
    \big|E&(A_{1}B_{2})+E(A_{3}B_{2})\big|
    +\big|E(A_{3}B_{4})+E(A_{5}B_{4})\big|+\cdots\\
    &+\big|E(A_{2N-3}B_{2N-2})+E(A_{2N-1}B_{2N-2})\big|\\
    &+\big|E(A_{2N-1}B_{2N})-E(A_{1}B_{2N})\big| \\
    \le & 2N-2.
  \end{split}
  \label{eq:2}
\end{equation}
This inequality applies on the ensemble on which all experimental
setups would give results, i.e., $A_{j},B_{k}\neq 0, \forall\,j,k$.
We would like an inequality that applies on correlations we can
obtain from experiment, such as $E(A_1B_2|\Lambda_{A_1B_2})$. To do
this, we note that the above inequality can be written
\begin{equation}
  \begin{split}
    \big|E&(A_{1}B_{2}|\Lambda_0)+E(A_{3}B_{2}|\Lambda_0)\big|
    +\big|E(A_{3}B_{4}|\Lambda_0)+E(A_{5}B_{4}|\Lambda_0)\big|\\
    &\phantom{\le}+\cdots+\big|E(A_{2N-3}B_{2N-2}|\Lambda_0)+E(A_{2N-1}B_{2N-2}|\Lambda_0)\big|\\
    &\phantom{\le}+\big|E(A_{2N-1}B_{2N}|\Lambda_0)-E(A_{1}B_{2N}|\Lambda_0)\big|\\
    \le &2N-2,
  \end{split}
\label{eq:3}
\end{equation}
where $\Lambda_0=\Lambda_{A_{1}B_{2}A_{3}B_{4}\ldots A_{2N-1}B_{2N}}$
denotes the ensemble where all measurements give results. Since
$E(A_{j}B_{k}|\Lambda_0)$ are not experimentally accessible, we need to
relate the ensemble $\Lambda_0$ to the ensembles $\Lambda_{A_jB_k}$, and
we do that by formally defining
\begin{equation}
\delta_{2N,2}=\min_{\text{settings}}P(\Lambda_0|\Lambda_{A_{j}B_{k}}).
\label{eq:6}
\end{equation}
We arrive at the following result:

\emph{Lemma 1.} Relation (\ref{eq:6}) between the subensemble that
obeys the BC inequality and the subensemble we see in experiment
enables the inequality
\begin{equation}
\begin{split}
    \big|E&(A_{1}B_{2}|\Lambda_{A_{1}B_{2}})+E(A_{3}B_{2}|\Lambda_{A_{3}B_{2}})\big|+\big|E(A_{3}B_{4}|\Lambda_{A_{3}B_{4}})\\
    &\phantom{\le}+E(A_{5}B_{4}|\Lambda_{A_{5}B_{4}})\big|+\cdots+\big|E(A_{2N-1}B_{2N}|\Lambda_{A_{2N-1}B_{2N}})\\
    &\phantom{\le}-E(A_{1}B_{2N}|\Lambda_{A_{1}B_{2N}})\big|\\
    & \le 2N-2\delta_{2N,2}.
\end{split}
\end{equation}

\emph{Proof.} Clearly, $\Lambda_0\subset\Lambda_{A_{j}B_{k}}$, so we
can split $\Lambda_{A_{j}B_{k}}$ into two subensembles $\Lambda_0$
(where all measurement settings give detections), and
$\Lambda_*=\Lambda_{A_jB_k}\setminus\Lambda_0$ (where $A_jB_k$ give
detections but one or more of the others do not). Note that
$\Lambda_0\cup\Lambda_*=\Lambda_{A_jB_k}$. We can write
\begin{equation}
\begin{split}
\Big|E&(A_jB_k|\Lambda_{A_jB_k})-\delta_{2N,2} E(A_jB_k|\Lambda_0)\Big|\\
&\le\Big|P(\Lambda_*|\Lambda_{A_jB_k})E(A_jB_k|\Lambda_*)\Big| +\Big|P(\Lambda_0|\Lambda_{A_jB_k})E(A_jB_k|\Lambda_0)\\
&\quad
-\delta_{2N,2} E(A_jB_k|\Lambda_0)\Big|\\
&=P(\Lambda_*|\Lambda_{A_jB_k})\Big|E(A_jB_k|\Lambda_*)\Big|\\
&\quad +\Big[P(\Lambda_0|\Lambda_{A_jB_k})
-\delta_{2N,2}\Big]\Big|E(A_jB_k|\Lambda_0)\Big|\\
&\le P(\Lambda_*|\Lambda_{A_jB_k})E\Big(|A_jB_k|\Big|\Lambda_*\Big)\\
&\quad +\Big[P(\Lambda_0|\Lambda_{A_jB_k})
-\delta_{2N,2}\Big]E\Big(|A_jB_k|\Big|\Lambda_0\Big)\\
&=1-\delta_{2N,2},
\end{split}
\end{equation}
Now,
\begin{equation}
\begin{split}
  \big|E&(A_iB_k|\Lambda_{A_iB_k})\pm E(A_jB_k|\Lambda_{A_jB_k})\big|\\
  &\le\delta_{2N,2}\big|E(A_iB_k|\Lambda_0)\pm E(A_jB_k|\Lambda_0)\big|+\big|E(A_iB_k|\Lambda_{A_iB_k})\\
  &\phantom{\le}-
  \delta_{2N,2}E(A_iB_k|\Lambda_0)\big|+\big|E(A_jB_k|\Lambda_{A_jB_k})-
  \delta_{2N,2}E(A_jB_k|\Lambda_0)\big|\\
  &\le\delta_{2N,2}\big|E(A_iB_k|\Lambda_0)\pm
  E(A_jB_k|\Lambda_0)\big| +2-2\delta_{2N,2}, \label{eq:4}
\end{split}
\end{equation}
so that finally,
\begin{equation}
\begin{split}
    &\big|E(A_{1}B_{2}|\Lambda_{A_{1}B_{2}})+E(A_{3}B_{2}|\Lambda_{A_{3}B_{2}})\big|+\big|E(A_{3}B_{4}|\Lambda_{A_{3}B_{4}})\\
    &\phantom{\le}+E(A_{5}B_{4}|\Lambda_{A_{5}B_{4}})\big|+\cdots+\big|E(A_{2N-1}B_{2N}|\Lambda_{A_{2N-1}B_{2N}})\\
    &\phantom{\le}-E(A_{1}B_{2N}|\Lambda_{A_{1}B_{2N}})\big|\\
    & \le
    \delta_{2N,2}(2N-2)+N(2-2\delta_{2N,2})\\
    &= 2N-2\delta_{2N,2}. \label{eq:5}
\end{split}
\end{equation}
\hfill\endproof

The following lemma gives the relation between the constant
$\delta_{2N,2}$ and the efficiency in the symmetric case.

\emph{Lemma 2.} In the symmetric case,
\begin{equation}
  \label{eq:9}
  \delta_{2N,2}\ge2N-1-\frac{2N-2}{\eta}.
\end{equation}
\emph{Proof.} We have
\begin{equation}
\begin{split}
&P(\Lambda_{A_{3}}|\Lambda_{A_{1}B_{2}})\\
&=\frac{P(\Lambda_{A_{1}A_{3}}|\Lambda_{B_{2}})}{P(\Lambda_{A_{1}}|\Lambda_{B_{2}})}
\\
&=\frac{P(\Lambda_{A_{1}}|\Lambda_{B_{2}})+P(\Lambda_{A_{3}}|\Lambda_{B_{2}})
-P(\Lambda_{A_{1}}\cup\Lambda_{A_{3}}|\Lambda_{B_{2}})}
{P(\Lambda_{A_{1}}|\Lambda_{B_{2}})}
\\
&\ge 1+\frac{\eta-1}{P(\Lambda_{A_{1}}|\Lambda_{B_{2}})}
\\
&\ge 2-\frac{1}{\eta},
\end{split}
\end{equation}
which gives
\begin{equation}
  \begin{split}
    \label{eq:7}
    P(\Lambda_{A_3B_4}|\Lambda_{A_1B_2})&=P(\Lambda_{A_3}|\Lambda_{A_1B_2})
    +P(\Lambda_{B_4}|\Lambda_{A_1B_2})\\
    &\phantom=
    -P(\Lambda_{A_3}\cup\Lambda_{B_4}|\Lambda_{A_1B_2})
    \\
    &\ge 2\Big(2-\frac{1}{\eta}\Big)-1\\
    &\
    = 3-\frac{2}{\eta}.
  \end{split}
\end{equation}
Now (Bonferroni),
\begin{equation}
  \begin{split}
    \label{eq:8}
    P(\Lambda_0|\Lambda_{A_{1}B_{2}})&=P(\Lambda_{A_{3}B_{4}}\cap\Lambda_{A_{3}B_{4}}\cap\cdots\cap
    \Lambda_{A_{2N-1}B_{2N}}|\Lambda_{A_{1}B_{2}})
    \\
    &\ge P(\Lambda_{A_{3}B_{4}}|\Lambda_{A_{1}B_{2}})+P(\Lambda_{A_{3}B_{4}}|\Lambda_{A_{1}B_{2}})+\cdots
    \\
    &\phantom{\ge}+P(\Lambda_{A_{2N-1}B_{2N}}|\Lambda_{A_{1}B_{2}})-(N-2)
    \\
    &\ge (N-1)\left(3-\frac2{\eta}\right)-(N-2)
    \\
    &=2N-1-\frac{2N-2}{\eta}.
  \end{split}
\end{equation}
Taking the minimum over the possible measurement settings immediately
gives the lemma.\hfill\endproof

These two lemmas give the BC inequality for the symmetric case as
\begin{equation}
  \begin{split}
    &\big|E(A_{1}B_{2}|\Lambda_{A_{1}B_{2}})+E(A_{3}B_{2}|\Lambda_{A_{3}B_{2}})\big|+\big|E(A_{3}B_{4}|\Lambda_{A_{3}B_{4}})
    \\
    &\phantom{\le}+E(A_{5}B_{4}|\Lambda_{A_{5}B_{4}})\big|
    +\cdots+\big|E(A_{2N-1}B_{2N}|\Lambda_{A_{2N-1}B_{2N}})
    \\
    &\phantom{\le}-E(A_{1}B_{2N}|\Lambda_{A_{1}B_{2N}})\big|
    \\
    &\le 2N-2\left(2N-1-\frac{2N-2}{\eta}\right)
    \\
    &=2(N-1)\left(\frac{2}{\eta}-1\right).
  \end{split}
  \label{eq:10}
\end{equation}
For a generic value $\beta$ on the left-hand side,
\begin{equation}
\beta\le2(N-1)\left(\frac{2}{\eta}-1\right),
\end{equation}
which leads to
\begin{equation}
\eta\leq \frac{2(N-1)}{N-1+\frac{\beta}{2}}.
\label{heta_beta_sim_ineq}
\end{equation}
Inserting the value of $\beta=2N\cos(\pi/2N)$ predicted by QM, we
arrive at the right-hand side of Eq.~(\ref{heta_quantum_sim}).

%%%%%%%%%%%%%%%%%%%%%%%%%%%%%%%%%%%%%%%%%%%%%%%%%%

The relation between the constant $\delta_{2N,2}$ and the efficiency
in the asymmetric case is as follows.

\emph{Lemma 3.} In the asymmetric case,
\begin{equation}
  \label{eq:12}
  \delta_{2N,2}\ge2N-1-\frac{N-1}{\eta_{A}}-\frac{N-1}{\eta_{B}},
\end{equation}

\emph{Proof.} The above approach gives
\begin{equation}
  \label{eq:11}
  \begin{split}
    P(\Lambda_{A_{3}B_{4}}|\Lambda_{A_{1}B_{2}})&=P(\Lambda_{A_{3}}|\Lambda_{A_{1}B_{2}})
    +P(\Lambda_{B_{4}}|\Lambda_{A_{1}B_{2}})\\
    &\phantom=-P(\Lambda_{A_{3}}\cup\Lambda_{B_{4}}|\Lambda_{A_{1}B_{2}})\\
    &\ge 2-\frac{1}{\eta_{A}}+2-\frac{1}{\eta_{B}}-1\\
    &=3-\frac{1}{\eta_{A}}-\frac{1}{\eta_{B}}.
  \end{split}
\end{equation}
The proof proceeds as that of Lemma 2.\hfill\endproof

Lemma 1 and Lemma 3 give the BC inequality for the asymmetric case
as
\begin{equation}
  \begin{split}
    &\big|E(A_{1}B_{2}|\Lambda_{A_{1}B_{2}})+E(A_{3}B_{2}|\Lambda_{A_{3}B_{2}})\big|+\big|E(A_{3}B_{4}|\Lambda_{A_{3}B_{4}})\\
    &\phantom{\le}+E(A_{5}B_{4}|\Lambda_{A_{5}B_{4}})\big|+\cdots+\big|E(A_{2N-1}B_{2N}|\Lambda_{A_{2N-1}B_{2N}})\\
    &\phantom{\le}-E(A_{1}B_{2N}|\Lambda_{A_{1}B_{2N}})\big|\\
    & \le 2(N-1)\left(\frac{1}{\eta_{A}}+\frac{1}{\eta_{B}}-1\right),
  \end{split}
  \label{eq:13}
\end{equation}
and, as before, for a value $\beta$ on the left-hand side
\begin{equation}
\beta
\le2(N-1)\left(\frac{1}{\eta_{A}}+\frac{1}{\eta_{B}}-1\right),
\end{equation}
or, equivalently,
\begin{equation}
\eta_{A}\le
\frac{1}{\frac{\beta}{2(N-1)}+1
-\frac1{\eta_{B}}}.
\label{heta_beta_asim_ineq}
\end{equation}
Again, for the quantum prediction on $\beta$ we obtain the
right-hand side of Eq.~(\ref{heta_quantum_asim}).

A particularly interesting case is when $\eta_{B}=1$. In terms of
a generic $\beta$ we have
\begin{equation}
  \label{}
  \eta_{A}\le
  \frac{2(N-1)}{\beta}.
\end{equation}
and, in particular for $\beta=2N\cos(\pi/2N)$,
\begin{equation}
  \eta_{A}\le
  \frac{N-1}{N\cos\left(\frac{\pi}{2N}\right)}.
\end{equation}

%%%%%%%%%%%%%%%%%%%%%%%%%%%%%%%%%%%%%%%%%%%%%%%%%%%%%%%%%%%%%%%%%%%

\subsection{\label{2D}Sufficient condition for symmetric efficiencies}

%%%%%%%%%%%%%%%%%%%%%%%%%%%%%%%%%%%%%%%%%%%%%%%%%%%%%%%%%%%%%%%%%%%

To prove sufficiency of the established bounds, it is convenient to go
back to our first approach to an LHV model, in terms of ensembles of
pairs of particles, with pairs of specified values occurring at a
given probability. We will simply build an LHV model with the desired
$\beta$ and $\eta$.

We start by splitting the total ensemble into subensembles
$\Lambda_i$ that collect states that have exactly $i$ non-detections
(zero values) of the constituent $A_j$'s and $B_k$'s. We note that
the $\Lambda_0$ so defined coincides with the $\Lambda_0$ defined at
inequality (\ref{eq:3}), and therefore that the BC inequality holds
for it. In fact, we have the following lemma:

\emph{Lemma 4.} It is possible to construct a LHV model so that the
results from the subensemble $\Lambda_0$ satisfy
$E(A_j|\Lambda_0)=E(B_k|\Lambda_0)=0$ and saturate the BC
inequality.

\emph{Proof.} Let $\Lambda_0$ consist of $4N$ states
($n=1,\ldots,2N$ and $m=\pm1$), all with equal probability, defined
so that
\begin{equation}
  \label{eq:14}
  A_j \text{ and }B_j =
  \begin{cases}
    \phantom{-}m, & j<n\\
    -m, & j\ge n
  \end{cases}
\end{equation}
It is immediately obvious that the individual results have equal
probability, and it is simple to verify that
\begin{equation}
  \begin{split}
    E(A_1B_2|\Lambda_0)&=E(A_3B_2|\Lambda_0)\\
    &=E(A_3B_4|\Lambda_0)\\
    &=\cdots = E(A_{2N-1}B_{2N}|\Lambda_0)\\
    &=-E(A_1B_{2N}|\Lambda_0)\\
    &=1-\tfrac1N.
    \label{eq:15}
  \end{split}
\end{equation}
Thus, the BC inequality is saturated by this model. \hfill\endproof

The subensembles where one or more non-detections occur are not
hindered by the BC inequality. Indeed, for those that give
well-defined correlations we have the following result:

\emph{Lemma 5.} It is possible to construct a LHV model so that the
results from the subensembles $\Lambda_i$, $1\le{}i\le{}2N-2$,
satisfy $E(A_j|\Lambda_i)=E(B_k|\Lambda_i)=0$ and give all
correlations the extreme value 1, and therefore maximally violate
the BC inequality.

\emph{Proof.} Let $\Lambda_1$ consist of $4N$ states
($n=1,\ldots,2N$ and $m=\pm1$), all with equal probability, defined
so that
\begin{equation}
  \label{eq:16}
  A_j \text{ and }B_j =
  \begin{cases}
    \phantom{-}m, & j<n\\
    \phantom{-}\,0, & j=n\\
    -m, & j>n.
  \end{cases}
\end{equation}
It is again immediately obvious that the individual results have equal
probability, and this time it is also obvious that
\begin{equation}
  \begin{split}
    E(A_1B_2|\Lambda_1)&=E(A_3B_2|\Lambda_1)\\
    &=E(A_3B_4|\Lambda_1)\\
    &=\cdots = E(A_{2N-1}B_{2N}|\Lambda_1)\\
    &=-E(A_1B_{2N}|\Lambda_1)\\
    &=1.
    \label{eq:17}
  \end{split}
\end{equation}
Ensembles $\Lambda_i$ with this property for $i>1$ can trivially be
constructed from $\Lambda_1$ by adding events with additional zeros
and thus, the lemma holds for those as well. \hfill\endproof

We are now in a suitable position to build an LHV model for the
required values of $\eta$ and $\beta$. The existence is sufficiently
important to give the result the status of a theorem.

{\em Theorem 2.} Sufficient condition for $\eta_{A}=\eta_{B}=\eta$:
When $2N-2 \leq \beta \leq 2N$ we can always build an LHV model with
\begin{equation}
\eta= \frac{2(N-1)}{N-1+\frac{\beta}{2}}.
\label{heta_beta_sim}
\end{equation}

{\em Proof.} We use the above ensemble construction of $\Lambda_0$
and $\Lambda_1$, and also a subensemble with no detections
$\Lambda_{2N}$; we let the other subensembles have probability zero.
In this model, under the assumption of independent errors, the
probabilities of single detection and coincidence are
\begin{subequations}
  \label{eq:18}
   \begin{align}
    P(\Lambda_0)+\left(1-\tfrac1{2N}\right)P(\Lambda_1)&=\eta,\\
    P(\Lambda_0)+\left(1-\tfrac1N\right)P(\Lambda_1)&=\eta^2.
  \end{align}
\end{subequations}
Solving for the unknown probabilities, we obtain
\begin{subequations}
  \begin{align}
    P(\Lambda_0)&=(2N-1)\eta^2-(2N-2)\eta,\\
    P(\Lambda_1)&=2N(\eta-\eta^2).
    \label{eq:19}
  \end{align}
\end{subequations}
We also obtain
\begin{equation}
  \label{eq:20}
  \begin{split}
    E(&A_1B_2|\Lambda_{A_1B_2})=\cdots = E(A_{2N-1}B_{2N}|\Lambda_{A_{2N-1}B_{2N}})\\
    &=-E(A_1B_{2N}|\Lambda_{A_1B_{2N}})\\
    &=\frac{\left(1-\tfrac1N\right)P(\Lambda_0)+\left(1-\tfrac1N\right)P(\Lambda_1)}
    {P(\Lambda_0)+\left(1-\tfrac1N\right)P(\Lambda_1)}\\
    &=\left(1-\frac1N\right)\frac{2\eta-\eta^2}{\eta^2}\\
    &=\left(1-\frac1N\right)\left(\frac2{\eta}-1\right).
  \end{split}
\end{equation}
This makes the left-hand side of the BC inequality obey
\begin{equation}
\begin{split}
    &\big|E(A_{1}B_{2}|\Lambda_{A_{1}B_{2}})+E(A_{3}B_{2}|\Lambda_{A_{3}B_{2}})\big|+\big|E(A_{3}B_{4}|\Lambda_{A_{3}B_{4}})\\
    &\phantom{\le}+E(A_{5}B_{4}|\Lambda_{A_{5}B_{4}})\big|+\cdots+\big|E(A_{2N-1}B_{2N}|\Lambda_{A_{2N-1}B_{2N}})\\
    &\phantom{\le}-E(A_{1}B_{2N}|\Lambda_{A_{1}B_{2N}})\big|\\
    & =(2N-2)\left(\frac2{\eta}-1\right)\\
    & =\beta.
\end{split}
\end{equation}
Solving for $\eta$, we arrive at Eq.~(\ref{heta_beta_sim}).
\hfill\endproof

%%%%%%%%%%%%%%%%%%%%%%%%%%%%%%%%%%%%%%%%%%%%%%%%%%%%%%%%%%%%%%%%%%%

\subsection{\label{2E}Sufficient condition for the asymmetric case}

%%%%%%%%%%%%%%%%%%%%%%%%%%%%%%%%%%%%%%%%%%%%%%%%%%%%%%%%%%%%%%%%%%%

To complete the sufficiency proof for $\eta_{A}\neq\eta_{B}$, we
first need to redefine our subensembles, to reflect the asymmetry of
the two detectors. Here, we split the total ensemble into
subensembles $\Lambda_{i,l}$ that collect states that have exactly
$i$ non-detections (zero values) of the constituent $A_j$'s and
exactly $l$ non-detections of the constituent $B_k$'s. We note that
again the $\Lambda_{0,0}$ so defined coincides with the $\Lambda_0$
defined at inequality (\ref{eq:3}), the BC inequality holds for it,
and Lemma 4 gives a LHV model that saturates the BC inequality.

The subensembles where one or more non-detections occur are still
not hindered by the BC inequality. Indeed, for those that give
well-defined correlations we have the following result:

\emph{Lemma 6.} It is possible to construct a LHV model so that the
results from the subensembles $\Lambda_{i,l}$, $0\le{}i,l\le{}N-1$
and not both zero, satisfy
$E(A_j|\Lambda_{i,l})=E(B_k|\Lambda_{i,l})=0$ and give all
correlations the extreme value 1, and therefore maximally violate
the BC inequality.

\emph{Proof.} In the case $l=0$, let $\Lambda_{1,0}$ consist of $2N$
states ($n=1,\ldots,N$ and $m=\pm1$), all with equal probability,
defined so that
\begin{equation}
\label{eq:21}
  A_j=
  \begin{cases}
    \phantom{-}m, & j<2n-1\\
    \phantom{-}\,0, & j=2n-1\\
    -m, & j>2n-1
  \end{cases}
  \text{ and }B_j =
  \begin{cases}
    \phantom{-}m, & j<2n\\
    -m, & j\ge 2n.
  \end{cases}
\end{equation}
Once more, it is immediately obvious that the individual results have
equal probability; it is also obvious that
\begin{equation}
  \begin{split}
    E(A_1B_2|\Lambda_{1,0})&=E(A_3B_2|\Lambda_{1,0})\\
   &=\cdots = E(A_{2N-1}B_{2N}|\Lambda_{1,0})\\
   &=-E(A_1B_{2N}|\Lambda_{1,0})=1.
    \label{eq:22}
  \end{split}
\end{equation}
The case $i=0$ is handled similarly, and the cases when both $i$ and
$l$ are nonzero can trivially be constructed by adding events with
additional zeros to, say, $\Lambda_{1,0}$ and thus, the lemma holds
for these cases as well. \hfill\endproof

We are now in a suitable position to build an LHV model for the
required values of $\eta$ and $\beta$.

{\em Theorem 3.} Sufficient condition for $\eta_{A}\neq\eta_{B}$:
When $2N-2 \leq \beta \leq 2N$ we can always build an LHV model with
\begin{equation}
\eta_{A}=
\frac{1}{\frac{\beta}{2(N-1)}+1
-\frac1{\eta_{B}}}.
\label{eq:23}
\end{equation}

{\em Proof.} We use the above ensemble construction of
$\Lambda_{0,0}$, $\Lambda_{1,0}$, and $\Lambda_{0,1}$, and also a
subensemble with no detections $\Lambda_{N,N}$; we let the other
subensembles have probability zero. In this model, under the
assumption of independent errors, the probabilities of single
detection and coincidence are
\begin{subequations}
  \label{eq:24}
   \begin{align}
     P(\Lambda_{0,0})+P(\Lambda_{0,1})
     +\left(1-\tfrac1N\right)P(\Lambda_{1,0})&=\eta_A,\\
     P(\Lambda_{0,0})+\left(1-\tfrac1N\right)P(\Lambda_{0,1})
     +P(\Lambda_{1,0})&=\eta_B,\\
     P(\Lambda_{0,0})+\left(1-\tfrac1N\right)\left[P(\Lambda_{0,1})+P(\Lambda_{1,0})\right]&=\eta_A\eta_B.
  \end{align}
\end{subequations}
Solving for the unknown probabilities, we obtain
\begin{subequations}
  \begin{align}
    P(\Lambda_{0,0})&=(2N-1)\eta_A\eta_B-(N-1)(\eta_A+\eta_B),\\
    P(\Lambda_{0,1})&=N(\eta_A-\eta_A\eta_B),\\
    P(\Lambda_{1,0})&=N(\eta_B-\eta_A\eta_B).
    \label{eq:25}
  \end{align}
\end{subequations}
We also obtain
\begin{equation}
  \label{eq:26}
  \begin{split}
    E(&A_1B_2|\Lambda_{A_1B_2})= \cdots = E(A_{2N-1}B_{2N}|\Lambda_{A_{2N-1}B_{2N}})\\
    &=-E(A_1B_{2N}|\Lambda_{A_1B_{2N}})\\
    &=\frac{\left(1-\tfrac1N\right)
      \left[P(\Lambda_{0,0})
      +P(\Lambda_{0,1})+P(\Lambda_{1,0})\right]}
    {P(\Lambda_{0,0})
      +\left(1-\tfrac1N\right)\left[
      P(\Lambda_{0,1})+P(\Lambda_{1,0})\right]}\\
    &=\left(1-\frac1N\right)\frac{\eta_A+\eta_B-\eta_A\eta_B}{\eta_A\eta_B}\\
    &=\left(1-\frac1N\right)\left(\frac1{\eta_A}+\frac1{\eta_B}-1\right).
  \end{split}
\end{equation}
This makes the left-hand side of the BC inequality obey
\begin{equation}
\begin{split}
    &\big|E(A_{1}B_{2}|\Lambda_{A_{1}B_{2}})+E(A_{3}B_{2}|\Lambda_{A_{3}B_{2}})\big|+\big|E(A_{3}B_{4}|\Lambda_{A_{3}B_{4}})\\
    &\phantom{\le}+E(A_{5}B_{4}|\Lambda_{A_{5}B_{4}})\big|+\cdots+\big|E(A_{2N-1}B_{2N}|\Lambda_{A_{2N-1}B_{2N}})\\
    &\phantom{\le}-E(A_{1}B_{2N}|\Lambda_{A_{1}B_{2N}})\big|\\
    & =(2N-2)\left(\frac1{\eta_A}+\frac1{\eta_B}-1\right)\\
    & =\beta.
\end{split}
\end{equation}
Solving for $\eta_A$, we arrive at Eq.~(\ref{eq:23}).
\hfill\endproof

%%%%%%%%%%%%%%%%%%%%%%%%%%%%%%%%%%%%%%%%%%%%%%%%%%%%%%%%%%%%%%%%%%%

\section{\label{3}Conclusions}

%%%%%%%%%%%%%%%%%%%%%%%%%%%%%%%%%%%%%%%%%%%%%%%%%%%%%%%%%%%%%%%%%%%

We have obtained the minimum detection efficiency required for a
loophole-free violation of the BC chained Bell inequalities
involving $N$ settings per observer. If both particles are detected
with the same efficiency, the minimum detection efficiency is given
by Eq.~(\ref{heta_quantum_sim}). If the particles are detected with
different efficiencies the minimum efficiencies are related by
Eq.~(\ref{heta_quantum_asim}).

The required detection efficiency {\em increases} with the number of
settings and tends to one as $N$ tends to infinity. This result
shows that the BC inequalities are not adequate for closing the
detection loophole. At this point, one should note that the BC
inequalities are useful in other situations, where other properties
than a high detection bound are important \cite{AKLZ99, Peres00,
BHK05, BKP06, CR08}.

Our results also establish the connection between the amount of
violation ${\cal{D}}$ and $\eta_{\rm crit}$ for the BC inequalities.
From Eqs.~(\ref{DBC}) and (\ref{heta_quantum_sim}), if
$\eta_{A}=\eta_{B}$, we obtain
\begin{equation}
\eta_{\rm crit}=\frac{2}{{\cal D} + 1},
\end{equation}
which establishes a close relation between $\eta_{\rm crit}$ and
${\cal D}$, similar to the one already found for the Mermin Bell
inequalities \cite{CRV08}. If $\eta_{A}\neq\eta_{B}$, from
Eq.~(\ref{heta_quantum_asim}), we have
\begin{equation}
\eta_{A\;{\rm crit}} = \frac{1}{{\cal D}+1-\frac{1}{\eta_{B\;{\rm
crit}}}}.
\end{equation}

Notice that ${\cal D}$ is related to the minimum visibility ${\cal
V}_{\rm crit}$ required to violate the BC chained Bell inequalities,
when, instead of $|\psi^-\rangle$, we have $\rho = {\cal V}
\,|\psi^-\rangle \langle \psi^-|+ (1-{\cal V}) \openone /4$, where
$\openone$ is the identity matrix. Specifically, a simple
calculation shows that ${\cal V}_{\rm crit} = 1/{\cal D}$.
Curiously, the same relation between ${\cal V}_{\rm crit}$ and
${\cal D}$ is found in stabilizer Bell inequalities for graph states
\cite{CGR08}.

So far, we have assumed that the prepared states have perfect
visibility (${\cal V}=1$). The effect of a non-perfect visibility
${\cal V} < 1$ can be easily calculated by replacing
$\beta=2N\cos(\pi/2N)$ by $2{\cal V}N\cos(\pi/2N)$ in all the
previous results.

%%%%%%%%%%%%%%%%%%%%%%%%%%%%%%%%%%%%%%%%%%%%%%%%%%%%%%%%%%%%%%%%%%%

\section{Acknowledgments}

%%%%%%%%%%%%%%%%%%%%%%%%%%%%%%%%%%%%%%%%%%%%%%%%%%%%%%%%%%%%%%%%%%%

The authors thank M. Barbieri and I. Vi\-lla\-nue\-va for useful
conversations. A.C. and D.R. acknowledge support from the Spanish
MCI Project No. FIS2008-05596. A.C. acknowledges support from the
Junta de Andaluc\'{\i}a Excellence Project No. P06-FQM-02243. D.R.
acknowledges support from the Junta de Andaluc\'{\i}a Excellence
Project No. P07-FQM-03037.

%%%%%%%%%%%%%%%%%%%%%%%%%%% References %%%%%%%%%%%%%%%%%%%%%%%%%%%%

%%%%%%%%%%%%%%%%%%%%%%%%%%%%%%%%%%%%%%%%%%%%%%%%%%%%%%%%%%%%%%%%%%%


\begin{thebibliography}{99}

%%%%%%%%%%%%%%%%%%%%%%%%%%%%%%%%%%%%%%%%%%%%%%%%%%%%%%%%%%%%%%%%%%%

\bibitem{CHSH69}
J.F. Clauser, M.A. Horne, A. Shimony, and R.A. Holt,
%``Proposed experiment to test local hidden-variable theories'',
Phys. Rev. Lett. {\bf 23}, 880 (1969).

%%%%%%%%%%%%%%%%%%%%%%%%%%%%%%%%%%%%%%%%%%%%%%%%%%%%%%%%%%%%%%%%%%%%

\bibitem{Bell64}
J.S. Bell,
%``On the Einstein-Podolsky-Rosen paradox'',
Physics (Long Island City, N.Y.) {\bf 1}, 195 (1964).

%%%%%%%%%%%%%%%%%%%%%%%%%%%%%%%%%%%%%%%%%%%%%%%%%%%%%%%%%%%%%%%%%%%
% The detection loophole
%%%%%%%%%%%%%%%%%%%%%%%%%%%%%%%%%%%%%%%%%%%%%%%%%%%%%%%%%%%%%%%%%%%

\bibitem{Wigner70}
E.P. Wigner (private communication), see Ref. \cite{Pearle70}.

\bibitem{Pearle70}
P.M. Pearle,
%``Hidden-variable example based upon data rejection'',
Phys. Rev.~D {\bf 2}, 1418 (1970).

%%%%%%%%%%%%%%%%%%%%%%%%%%%%%%%%%%%%%%%%%%%%%%%%%%%%%%%%%%%%%%%%%%%
% Bell test without the detection loophole
%%%%%%%%%%%%%%%%%%%%%%%%%%%%%%%%%%%%%%%%%%%%%%%%%%%%%%%%%%%%%%%%%%%

\bibitem{RKVSIMW01}
M.A. Rowe, D. Kielpinski, V. Meyer, C. A. Sackett, W.M. Itano, C.
Monroe, and D.J. Wineland,
%``Experimental violation of a Bell's inequality
%with efficient detection'',
Nature (London) {\bf 409}, 791 (2001).

\bibitem{MMMOM08}
D.N. Matsukevich, P. Maunz, D. L. Moehring, S. Olmschenk, and C.
Monroe,
%``Bell inequality violation with two remote atomic qubits'',
Phys. Rev. Lett. {\bf 100}, 150404 (2008).

%%%%%%%%%%%%%%%%%%%%%%%%%%%%%%%%%%%%%%%%%%%%%%%%%%%%%%%%%%%%%%%%%%%
% Detection efficiency requirements for CHSH (I)
%%%%%%%%%%%%%%%%%%%%%%%%%%%%%%%%%%%%%%%%%%%%%%%%%%%%%%%%%%%%%%%%%%%

\bibitem{GM87}
A. Garg and N.D. Mermin,
%``Detector inefficiences
%in the Einstein-Podolsky-Rosen experiment'',
Phys. Rev.~D {\bf 35}, 3831 (1987).

%%%%%%%%%%%%%%%%%%%%%%%%%%%%%%%%%%%%%%%%%%%%%%%%%%%%%%%%%%%%%%%%%%%
% Detection efficiency requirements for CHSH (II)
%%%%%%%%%%%%%%%%%%%%%%%%%%%%%%%%%%%%%%%%%%%%%%%%%%%%%%%%%%%%%%%%%%%

\bibitem{Larsson98}
J.-\AA. Larsson,
%``Bell's inequality and detector inefficiency'',
Phys. Rev.~A {\bf 57}, 3304 (1998).

%%%%%%%%%%%%%%%%%%%%%%%%%%%%%%%%%%%%%%%%%%%%%%%%%%%%%%%%%%%%%%%%%%%
% Minimum detection efficiency for atom-photon Bell experiments
%%%%%%%%%%%%%%%%%%%%%%%%%%%%%%%%%%%%%%%%%%%%%%%%%%%%%%%%%%%%%%%%%%%

\bibitem{CL07}
A. Cabello and J.-\AA. Larsson,
%``Minimum detection efficiency for a loophole-free atom-photon Bell experiment'',
Phys. Rev. Lett. {\bf 98}, 220402 (2007).

\bibitem{BGSS07}
N. Brunner, N. Gisin, V. Scarani, and C. Simon,
%N. Brunner {\em et al.},
%``Detection loophole in asymmetric Bell experiments'',
%quant-ph/0702130.
Phys. Rev. Lett. {\bf 98}, 220403 (2007).

%%%%%%%%%%%%%%%%%%%%%%%%%%%%%%%%%%%%%%%%%%%%%%%%%%%%%%%%%%%%%%%%%%%

\bibitem{Mermin90}
N. D. Mermin,
%``Extreme quantum entanglement in a
%superposition of macroscopically distinct states'',
Phys. Rev. Lett. {\bf 65}, 1838 (1990).

%%%%%%%%%%%%%%%%%%%%%%%%%%%%%%%%%%%%%%%%%%%%%%%%%%%%%%%%%%%%%%%%%%%

\bibitem{CRV08}
A. Cabello, D. Rodr\'{\i}guez, and I. Villanueva,
%``Necessary and sufficient detection efficiency for the Mermin
%inequalities'',
Phys. Rev. Lett. {\bf 101}, 120402 (2008).

%%%%%%%%%%%%%%%%%%%%%%%%%%%%%%%%%%%%%%%%%%%%%%%%%%%%%%%%%%%%%%%%%%%

\bibitem{BC89}
S.L. Braunstein and C.M. Caves,
%``Chained Bell inequalities'',
in {\em Bell's Theorem, Quantum Theory, and Conceptions of the
Universe}, edited by M. Kafatos (Kluwer, Dordrecht, 1989), p.~27.

\bibitem{BC90}
S.L. Braunstein and C.M. Caves,
%``Wringing out better Bell inequalities'',
Ann. Phys. (N.Y.) {\bf 202}, 22 (1990).

%%%%%%%%%%%%%%%%%%%%%%%%%%%%%%%%%%%%%%%%%%%%%%%%%%%%%%%%%%%%%%%%%%%

\bibitem{Peres93}
A. Peres, {\em Quantum Theory: Concepts and Methods} (Kluwer,
Dordrecht, 1993).

%%%%%%%%%%%%%%%%%%%%%%%%%%%%%%%%%%%%%%%%%%%%%%%%%%%%%%%%%%%%%%%%%%%

\bibitem{Tsirelson80}
B. S. Cirel'son,
%``Quantum generalizations of Bell's inequality'',
Lett. Math. Phys. {\bf 4}, 93 (1980).

%%%%%%%%%%%%%%%%%%%%%%%%%%%%%%%%%%%%%%%%%%%%%%%%%%%%%%%%%%%%%%%%%%%

\bibitem{Wehner06}
S. Wehner,
%``Tsirelson bounds for generalized Clauser-Horne-Shimony-Holt
%inequalities'',
Phys. Rev.~A {\bf 73}, 022110 (2006).
%quant-ph/0510076.

%%%%%%%%%%%%%%%%%%%%%%%%%%%%%%%%%%%%%%%%%%%%%%%%%%%%%%%%%%%%%%%%%%%

\bibitem{BBDH97}
D. Boschi, S. Branca, F. De Martini, and L. Hardy,
%``Ladder proof of nonlocality without inequalities:
%Theoretical and experimental results'',
Phys. Rev. Lett. {\bf 79}, 2755 (1997).

\bibitem{BDDM05}
M. Barbieri, F. De Martini, G. Di Nepi, and P. Mataloni,
%`Towards a test of non-locality without ``supplementary assumptions''\,',
Phys. Lett.~A {\bf 334}, 23 (2005).

%%%%%%%%%%%%%%%%%%%%%%%%%%%%%%%%%%%%%%%%%%%%%%%%%%%%%%%%%%%%%%%%%%%

\bibitem{AKLZ99}
S. Aerts, P. Kwiat, J.-{\AA}. Larsson, and M. \.Zukowski,
%``Two-photon {F}ranson-type experiments and local realism.''
Phys. Rev. Lett. {\bf 83}, 2872 %--2875,
(1999).

\bibitem{Peres00}
A. Peres,
%``Bayesian analysis of Bell inequalities'',
Fortschr. Phys. {\bf 48}, 531 (2000).

\bibitem{BHK05}
J. Barrett, L. Hardy, and A. Kent,
%``No signaling and quantum key distribution'',
Phys. Rev. Lett. {\bf 95}, 010503 (2005).

\bibitem{BKP06}
J. Barrett, A. Kent, and S. Pironio,
%``Maximally nonlocal and monogamous quantum correlations'',
Phys. Rev. Lett. {\bf 97}, 170409 (2006).

\bibitem{CR08}
R. Colbeck and R. Renner,
%``Hidden variable models for quantum theory cannot have any local part'',
Phys. Rev. Lett. {\bf 101}, 050403 (2008).

%%%%%%%%%%%%%%%%%%%%%%%%%%%%%%%%%%%%%%%%%%%%%%%%%%%%%%%%%%%%%%%%%%%

\bibitem{CGR08}
A. Cabello, O. G{\"u}hne, and D. Rodr\'{\i}guez,
%``Mermin inequalities for perfect correlations'',
Phys. Rev. A {\bf 77}, 062106 (2008).

\end{thebibliography}
\end{document}